\documentclass[aps,prl,showpacs,superscriptaddress,reprint]{revtex4-1}

\usepackage{epsfig}
\usepackage{amsmath}
\usepackage{amssymb}
\usepackage{graphicx}
\usepackage{acronym}
\usepackage{color}
\usepackage[normalem]{ulem}

\begin{document}

\title{Specific heat anomaly in a supercooled liquid with amorphous
  boundary conditions (Includes SM)}

\author{Daniel A. M\'{a}rtin}

\affiliation{Instituto de Investigaciones Fisicoqu\'\i{}micas Te\'oricas y
  Aplicadas (INIFTA), CONICET and Facultad de Ciencias Exactas,
  Universidad Nacional de La Plata, c.c.~16, suc.~4, B1904DPI La
  Plata, Argentina}

\affiliation{Departamento de Ciencias B\'asicas, Facultad de Ingenier\'\i{}a, Universidad Nacional de La
  Plata, 1900 La Plata, Argentina}

\author{Andrea Cavagna}

\affiliation{Istituto Sistemi Complessi (ISC), Consiglio Nazionale
  delle Richerche (CNR), UOS Sapienza, Via dei Taurini, 19, 00185
  Roma, Italy}

\author{Tom\'as S. Grigera}

\thanks{Present address: Instituto de F\'\i{}sica de L\'\i{}quidos y
  Sistemas Biol\'ogicos (IFLYSIB), c.c. 565, 1900 La Plata, Argentina}

\affiliation{Instituto de Investigaciones Fisicoqu\'\i{}micas Te\'oricas y
  Aplicadas (INIFTA), CONICET and Facultad de Ciencias Exactas,
  Universidad Nacional de La Plata, c.c.~16, suc.~4, B1904DPI La
  Plata, Argentina}

\affiliation{CCT CONICET La Plata, Consejo Nacional de Investigaciones
  Cient\'\i{}ficas y T\'ecnicas, Argentina}

\affiliation{Departamento de F\'\i{}sica, Universidad Nacional de La
  Plata, c.c.\ 67, 1900 La Plata, Argentina}

\begin{abstract}

  We study the specific heat of a model
    supercooled liquid confined in a spherical cavity with amorphous
    boundary conditions.  We find the \emph{equilibrium} specific heat
    has a cavity-size-dependent peak as a function of temperature.
    The cavity allows us to perform a \ac{FSS} analysis, which
    indicates the peak persists at a finite temperature in the
    thermodynamic limit.  We attempt to collapse the data onto a
    \ac{FSS} curve according to different theoretical scenarii,
    obtaining reasonable results in two cases: a ``not-so-simple''
    liquid with nonstandard values of the exponents $\alpha$ and
    $\nu$, and \acf{RFOT}, with two different length scales.

\end{abstract}


\date{April 15, 2015}

\maketitle

\acrodef{ABC}[ABCs]{amorphous boundary conditions}
\acrodef{PBC}[PBCs]{periodic boundary conditions}
\acrodef{PTS}[PTS]{point-to-set}
\acrodef{RBC}[RBCs]{random boundary conditions}
\acrodef{MC}{Monte Carlo}
\acrodef{RFOT}{random first-order theory}
\acrodef{FSS}{finite-size scaling}
\acrodef{NSSL}{Not-so-simple liquid}

\def\Cv{C_V}
\def\Cvones{\Cv^{\text{(1S)}}}
\def\Cvvib{\Cv^{\text{(v)}}}
\def\Cvbulk{\Cv^{\text{(b)}}}
\def\Cvin{\Cv^{\text{(in)}}}
\def\Cvout{\Cv^{\text{(out)}}}
\def\pin{p^{\text{(in)}}}
\def\pout{p^{\text{(out)}}}

In fragile glassformers, the relaxation time increases faster than the
Arrhenius law as temperature is lowered \cite{review:angell95}.  This
implies that the effective barrier to relaxation grows on cooling,
which leads to expect concomitant structural, and perhaps
thermodynamic, changes.  Though not universally accepted, the idea
that a thermodynamic transition may underlie the dynamic glass
transition is old and at the core of \acf{RFOT} and other theoretical
approaches \cite{review:berthier2011}.  The question of the existence
of a transition is open; in fact \emph{structural} changes
accompanying the slowdown have been found only recently
\cite{self:nphys08, landscape:widmer-cooper08, glassthermo:lerner09,
  glassthermo:tanaka10, correlation-length:sausset11,
  glassthermo:coslovich2011, frustration:charbonneau12,
  correlation-length:hocky12, confinement:kob12}, after more than a
decade of study of dynamic correlations \cite{review:ediger00,
  review:sillescu99}.

The most general tools for probing structural correlations are the
``order-agnostic'' methods ---which include patch correlations
\cite{correlation-length:kurchan11, correlation-length:sausset11},
\acf{FSS} \cite{glassthermo:fernandez06,
  correlation-length:karmakar09}, \ac{PTS} \cite{dynamics:montanari06}
and its related correlations--- which do not need knowledge of the
order parameter.  Calculation of \ac{PTS} correlations involves the
study of confined systems, and in part for this reason a growing
number of studies of liquids under various confined geometries have
been reported, mainly cavities with \acl{ABC} (explained below)
\cite{mosaic:bouchaud04, mosaic:jack05, self:prl07, self:nphys08,
  confinement:berthier12, correlation-length:hocky12}, ``cavities''
with open directions 
\cite{confinement:berthier12, self:jcp13} and systems with pinned
particles \cite{confinement:berthier12, frustration:charbonneau12,
  confinement:karmakar13}.  These investigations have focused mostly
on density correlations, from which a correlation length can be
extracted.

Here we report numerical results on the specific heat $\Cv$ of a
system confined under \acf{ABC}, therefore combining the \ac{ABC} and
standard \ac{FSS} approaches \footnote[1]{Notice that one-time
  thermodynamic quantities, like energy, density or magnetization, are
  not sensitive to \ac{ABC} \cite{mosaic:zarinelli10,
    self:jstatmech10, confinement:krakoviack10}, while correlation
  functions and susceptibilities may be.}.  We find an anomalous peak
as a function of temperature. The algorithm we
  use (swap \ac{MC} \cite{self:pre01}) provides a complete sampling of
  configuration space at all the temperatures we report, so the peak
  is completely unrelated to the usual anomalies caused by the system
  falling out of equilibrium.  We use \ac{FSS} to study the changes of
  this \emph{thermodynamic} anomaly as the cavity is enlarged, and our
  results indicate that it remains at a finite temperature in the
  thermodynamic limit.  This is further evidence of the structural
  changes happening in supercooled liquids, and supports the existence
  of a thermodynamic transition.


We study the soft-sphere binary mixture of
ref.~\cite{soft-spheres:bernu87} with size ratio 1.2 and unit density.
To confine with \ac{ABC}, a spherical cavity of radius $R$ is created
in an equilibrium configuration from a \ac{PBC} system at temperature
$T$, introducing a hard wall that conserves density and composition
inside the cavity \cite{self:prl07, self:nphys08,
  self:jstatmech10}. Inside particles evolve with swap \ac{MC}
\cite{self:pre01} at the same temperature, while outside particles are
held fixed.  The specific heat is computed through energy
fluctuations,
$ \Cv= \overline{[\langle E^2\rangle -\langle E\rangle^2]}/(MT^2)$,
where $E$ is the energy, $M$ the number of cavity (free) particles,
and the overline means average over different realizations of the BCs.
All results correspond to the (meta)\emph{equilibrium supercooled
  liquid.}  We used the energy time correlation function (checking for
aging and finite-time effects) to estimate a correlation time and
ensure that all relevant quantities were computed using runs lasting
more than 100 relaxation times (including, self-consistently, the
energy correlation).  We used the bond orientation order parameter
$Q_6$ \cite{review:tanaka12} to exclude samples that showed signs of
crystallization and could give a spurious contribution to the liquid
$\Cv$.  Note that equilibration of small cavities is not problematic
since with swap Monte Carlo smaller cavities are faster (not slower)
than larger ones \cite{self:jcp12}.  For detailed description of
simulation and equilibration procedures and crystallization checks,
see SM \footnote[2]{See Supplemental Material [url], which includes Refs. \cite{R31,R32}.}.


\begin{figure}
\includegraphics[width=\columnwidth]{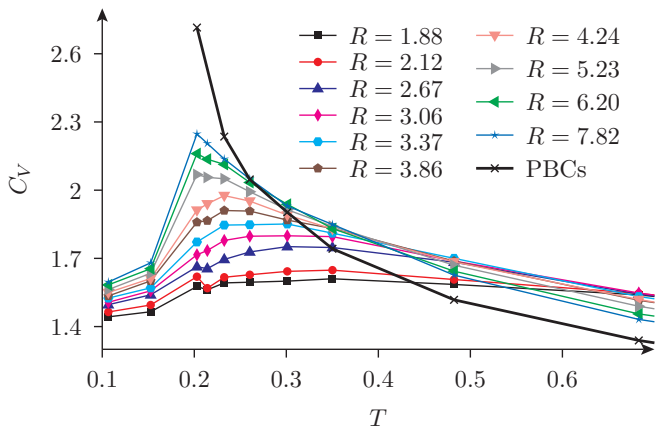}
 \caption{(color online) Specific heat vs.\ $T$ for \acf{ABC}.  Error bars omitted
   for clarity; the absolute error is bounded by 0.03.  \ac{PBC} data
   are for a system of $N=8192$ particles, where $\Cv$ can be measured
   down to $T\approx0.2$ (see SM for details).}
    \label{fig:Cv-T}
\end{figure}

Fig.~\ref{fig:Cv-T} shows the specific heat per mobile particle for
\ac{ABC} for several cavity sizes (from 28 to 2000 mobile particles),
displaying a peak.  The peak is not due to the system going out of
equilibrium, or to crystal formation.  For classical liquids, $\Cv$ is
expected to be monotonically decreasing with temperature (as has been
shown for a large number of liquid models
\cite{liquid-thermodynamics:rosenfeld98,
  liquid-thermodynamics:ingebrigtsen13} and follows from the phonon
theory of liquids \cite{liquid-thermodynamics:bolmatov12}): in this
sense, the observed peak is an anomaly.  A similar
  anomaly has been reported before \cite{self:pre01, Yan2004},
  although in rather small systems and without a \ac{FSS} analysis.

In the simplest scenario, the qualitative origin of this anomalous
peak can be explained if we accept that the effect of the border
penetrates into the cavity as far as a length-scale $\lambda(T)$, and
that this penetration length increases for lower $T$
\cite{self:prl07}.  At very high $T$ the effect of the boundary is
weak ($\lambda(T)$ is very small) and the $\Cv$ of the liquid inside
the cavity follows its bulk behavior (in our case well described at
high $T$ by the Rosenfeld-Tarazona law
$\Cv \sim \Cvbulk = A T^{-2/5}$).  For low $T$, on the other hand,
$\lambda(T)$ will be large compared to the size $R$, so that the
cavity will be almost frozen.  The crossover from an increase to
constant gives rise to the peak.  Notice we have assumed no particular
theory, nor any divergence of $\lambda(T)$ here, only that a very
small cavity (relative to $\lambda$) is stuck.  Hence, the mere
presence of this peak does not allow us to discriminate among
theoretical frameworks.  We need to be more quantitative.

The penetration length $\lambda$ is conceptually different from the
correlation length $\xi$ \cite{self:prl07, mosaic:zarinelli10,
  self:jcp13, arxiv:biroli14}.  The correlation length is a measure of
the distance that two points must be separated form each other so that
the local state is mutually independent (or in a cavity, how large
must the cavity be so that the state at the center is independent from
the state at the walls).  The penetration length is only meaningful in
the presence of domains, and is a measure of the width of the domain
walls (in a cavity, how far from the wall a point must be to be
independent from the state outside).  $\xi$ can also be
thought of as a measure of the size of the domains (or cooperatively
rearranging regions), and $\lambda$ as a measure of the interface
width, or the extent of the spatial fluctuations of the walls
separating such domains or regions.  The lengths can be coincide in
simple cases  (like the Ising model), but in principle they measure two
different phenomena.

There are thus two sides in our finite-size
  story: a finite size is needed for the \ac{ABC} border to have an
  effect, but the converse is not true: with \ac{PBC}, for instance,
  we can have finite size but no border effects.  Hence we must try to
  include, but separate, both effects: that of the \ac{ABC} border
  (related to $\lambda$) and that of the finite size (related to
  $\xi$).  We believe a reasonable way to do this, at least near the
  peak, is to write
\begin{equation}
\Cv(R,T)= R^{\alpha/\nu} \tilde c(y)  [1-  f(R/\lambda)].
\label{proposal}
\end{equation} 
The factor $R^{\alpha/\nu} \tilde c(y)$ is the border-free, \ac{FSS}
form of the specific heat, and it would be a safe bet in most
finite-size systems with \ac{PBC} \cite{book:newman99}.  Through the
scaling variable $y\equiv R^{1/\nu}(T-T_c)/T_c$, the finite-size term
contains all the information about the possible existence of a
finite-temperature transition, $T_c$, and about the correlation length
$\xi\sim (T-T_c)^{-\nu}$; hence $\tilde c(y)$ is a function of
$\xi/R$.  The second factor on the r.h.s.\ is meant to take care of
the border: for large $R/\lambda$ the border function $f\sim 0$ and
the effect of the border is negligible (but not necessarily that of
finite size).  But we need to be more specific as regards $f(x)$ to
test our scaling ansatz Eq.~\ref{proposal}, so we make the simplest
assumption: that the specific heat is zero exactly at the border and
relaxes exponentially to the (finite-size) \ac{PBC} value.  This
results in \cite{self:prl07}
\begin{equation}
f(x) =3\left[x^{-1}-2x^{-2}+2x^{-3}(1- e^{-x})\right],
\label{eqf}
\end{equation} 
with $x=R/\lambda$.

This form of $f(x)$ is certainly an approximation; to understand its
significance we must first make three remarks.  First,
Eq.~\ref{proposal} is qualitatively different from standard \ac{FSS}
only if $\lambda$ and $\xi$ are two different length scales, otherwise
the effect of the border is a mere decoration of the scaling function
and standard \ac{FSS} remains unchanged.  Second, in the critical
region, $R\sim\xi$, which follows because the peak position is given
by the position $y_0$ of the maximum of $\tilde c(y)$.  This follows
immediately in the usual FSS case; in the general case the analysis is
slightly more complicated (see SM \footnotemark[2]), but it remains
true that the peak position and its temperature shift are given by
$R^{1/\nu}(T-T_c)\simeq y_0$.  Third, (at least in the scenarios we
consider), $\lambda$ diverges at $T_c$, but not as fast as $\xi$.

When $R/\lambda\to\infty$, Eq.~\ref{eqf} gives correctly $f\to0$.
When $R/\lambda\to0$, $f(x)$ tends unphysically to 1, but in the
critical region $\lambda$ is large but $R\sim\xi$, so that
$R/\lambda\ll1$.  This means the $R/\lambda\to0$ limit of
$f(R/\lambda)$ is irrelevant in the critical region.  Very near $T_c$
though, (which is \emph{outside} the critical region at finite size)
our approximation will have the effect of making $\tilde c(y)$
divergent as $y\to0$.  Since this unwanted effect can be avoided at
the expense of introducing unknown parameters, we have preferred to
leave Eq.~\ref{eqf} as is.  See SM \footnotemark[2] for a discussion
of this point and the possible cure.

Finally, note that in Eq.~\ref{proposal} there are two different
mechanisms for the growth of the $\Cv$ peak as $R$ increases.  When
there are no border effects, the $1-f(R/\lambda)$ factor is absent,
and the growth of the peak is controlled by the $R^{\alpha/\nu}$
prefactor \cite{book:newman99}.  Hence, a nonzero $\alpha$ is normally
required to explain a growing (eventually diverging for $R\to\infty$)
peak.  When there is a border, then the last factor also produces a
(moderate) growth of the peak if $R$ grows faster than $\lambda$, so
that $\alpha = 0$ is compatible with a non-diverging growth of the
specific heat for $T\to T_c$.

With this in mind, we now scale our finite-size data according to
Eq.~\ref{proposal}, namely we try to collapse the data by plotting,
$C_V R^{-\alpha/\nu} [1- f(R/\lambda)]^{-1}$ vs.\
$R^{1/\nu}(T-T_c)/T_c$.  It is clearly useless to attempt to scale
with all $T_c$, $\nu$, $\alpha$ and $\lambda(T)$ free, as there are
too many parameters.  We will rather try to compare different
theoretical scenarios, thus fixing some of these parameters.

\begin{figure}
  \includegraphics[width=\columnwidth]{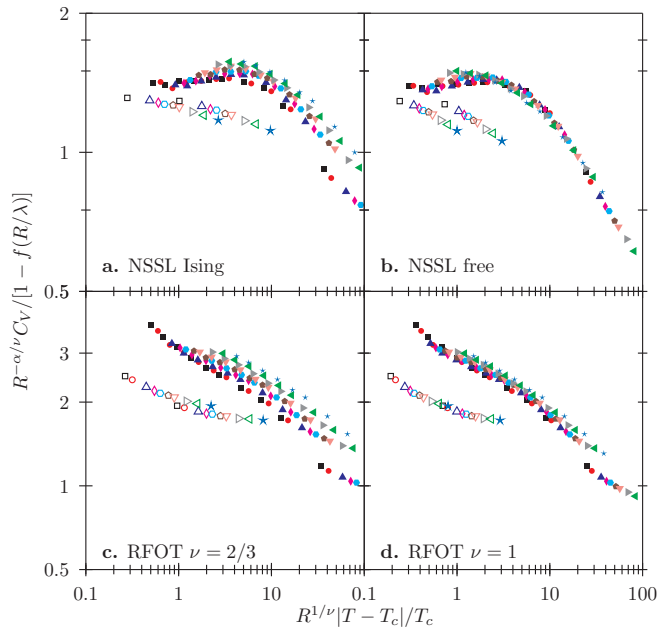}
  \caption{(color online) Attempts at scaling according to different
    theoretical scenarios.  {\bf a:} \ac{NSSL} with Ising exponents
    ($f(x)\equiv0$, $Tc=0.17$, $\alpha=0.11$, $\nu=0.63$).  {\bf b:}
    \ac{NSSL} with free exponents ($f(x)\equiv 0$, $Tc=0.175$,
    $\alpha=0.2$, $\nu=1.0$).  {\bf c:} RFOT with $\nu=2/d$ ($\alpha
    \equiv 0$, $Tc=0.17$, $\nu=2/3$, $f(x)$ as in Eq.~\ref{eqf}).
    {\bf d:} RFOT with $\nu=1$ ($\alpha \equiv 0$, $Tc=0.17$, $f(x)$
    as in Eq.~\ref{eqf}).  Colors and symbols indicate different radii
    (scheme as in Fig.~\ref{fig:Cv-T}), filled symbols are for
    $T>T_c$, open symbols for $T<T_c$.}
  \label{ABCfit}
\end{figure}

\paragraph{Simplest liquid.}  In the simplest possible physical
scenario we have no transition ($T_c=0$) and only one length scale,
$\lambda \sim \xi$ (i.e.\ $f\equiv0$).  In this case we are scaling
the data as $\Cv R^{-\alpha/\nu}$ vs $R^{1/\nu}T$.  We then need
$\alpha\neq 0$ to account for the growth of the peak.  Moreover, it
seems reasonable to assume that the standard RG scaling relation
(Josephson scaling) $\nu d=2-\alpha$ \cite{book:binney92} holds in
this simplest case.  We are therefore left with just one parameter,
$\alpha$.  The result is quite bad and no reasonable collapse is
obtained for any value of $\alpha$ (not shown).

\paragraph{\acf{NSSL}.}  What really seems to resist the scaling of
the data in the simplest case is the assumption $T_c=0$.  We therefore
relax this hypothesis, assuming that a standard phase transition
exists at a finite temperature (standard meaning there is only one
length scale, so that $\lambda \sim \xi$, and normal FSS ($f\equiv0$)
applies).  One such case is that invoked by Tanaka et
al.~\cite{glassthermo:tanaka10}, with Ising-like critical exponents
(which thus satisfy the relation $\nu d=2-\alpha$).  This proposal
does not achieve a reasonable collapse, irrespective of the value of
$T_c$ (Fig.~\ref{ABCfit}a).  Fern\'andez et
al.~\cite{glassthermo:fernandez06} have studied the specific heat of
our same system under \ac{PBC} and seemed to find a divergence at
temperature $T_c=0.195$, with $\alpha=0.9$.  With these values,
however, we fail to obtain a collapse (violating Josephson scaling
does not help much either).  If, however we leave \emph{all three}
parameters $\alpha$, $\nu$ and $T_c$ free, we get a reasonable
collapse for the data above $T_c$ (Fig.~\ref{ABCfit}b).

\paragraph{Mosaic liquid.}  Now we assume that the penetration length,
$\lambda$, and correlation length, $\xi$, grow differently. The
increase of the peak for increasing $R$ implies that $\lambda \ll
\xi$. This is the only case in which \ac{ABC} really have some
nontrivial qualitative effect, because this is the only way in which
we can achieve a growth of the peak with $\alpha=0$: in this case, the
specific heat has a kink in the bulk limit, rather than a divergence.
This is exactly what is supposed to happen in the \acf{RFOT}, as well
as in some mean-field spin-glass models, in particular the $p$-spin
\cite{mosaic:kirkpatrick87, p-spin:crisanti92, glassthermo:coluzzi99,
  mosaic:bouchaud04}.  The \ac{RFOT} transition is first order in the
sense that it has a discontinuous order parameter, but second order in
the Ehrenfest sense \cite{p-spin:crisanti92, glassthermo:coluzzi99}.
Quite generally $\alpha=0$ in RFOT, as a consequence of the fact that
the configurational entropy vanishes at $T_c$, giving a discontinuity
of the derivative of the total entropy at the transition
\cite{p-spin:crisanti92, glassthermo:coluzzi99, review:lubchenko07}.
There is theoretical \cite{arxiv:biroli14} as well as numerical
\cite{self:prl07, self:jcp13} evidence that within the RFOT scenario
indeed penetration and correlation length are different things and
that $\lambda \ll \xi$.

In this scenario both $\xi$ and $\lambda$ diverge at $T_c$, but with
different critical exponents.  For $\lambda(T)$, the prediction is
that $\lambda \sim \lvert T-T_c\rvert^{-1/2}$ in three dimensions
\cite{arxiv:biroli14}.  The exponent ruling the $\xi$ divergence is
$\nu=1/(d-\theta)$ \cite{mosaic:kirkpatrick89}, where $\theta$ is the
stiffness exponent, for which different values have been predicted.
Some approximations \cite{mosaic:kirkpatrick89} give $\theta=d/2$,
corresponding to $\nu=2/d$, while others \cite{nucleation:franz05,
  self:jcp13} give $\theta=d-1$, which yields instead $\nu=1$.  In
three dimensions, both predictions imply $\nu>1/2$ and thus
$\xi\gg\lambda$ near $T_c$.

To perform the $\lambda \neq \xi$ scaling we need to plot
$\Cv R^{-\alpha/\nu} [1- f(\lambda/R)]^{-1}$ vs.\
$R^{1/\nu}(T-T_c)/T_c$.  We use Eq.~\ref{eqf} as an approximation for
$f(x)$, but we need also $\lambda(T)$.  For this we have taken the
data of ref.~\cite{self:prl07} and fitted them to a power law
$a\lvert T-T_c\rvert^{1/2}$ \cite{arxiv:biroli14}, leaving $a$ as
fitting parameter but fixing $T_c$ self-consistently to the value that
gives the best collapse of $\Cv$.  The resulting scalings are shown in
Fig.~\ref{ABCfit} (panels c and d).  Both have $\alpha=0$ in
accordance to RFOT predictions, with $\lambda(T)$ for $T>T_c$ taken
from the power law fit as explained above.  $T_c$ is a free parameter,
while $\nu$ is fixed to the values $2/3$ and $1$ according to the
different predictions.  The \ac{RFOT} scaling with $\nu=2/3$
(Fig.~\ref{ABCfit}c) does not give a good collapse of the data, while
using $\nu=1$ does a rather good job (Fig.~\ref{ABCfit}d) for $T>T_c$.
For $T<T_c$ we do not have data to fit $\lambda(T)$, hence we have
used the same power law with a prefactor $a'$ chosen to give the best
collapse.  So the $T<T_c$ branch has a better-looking collapse but
with two free parameters instead of one.

\begin{figure}
 \includegraphics[width=\columnwidth]{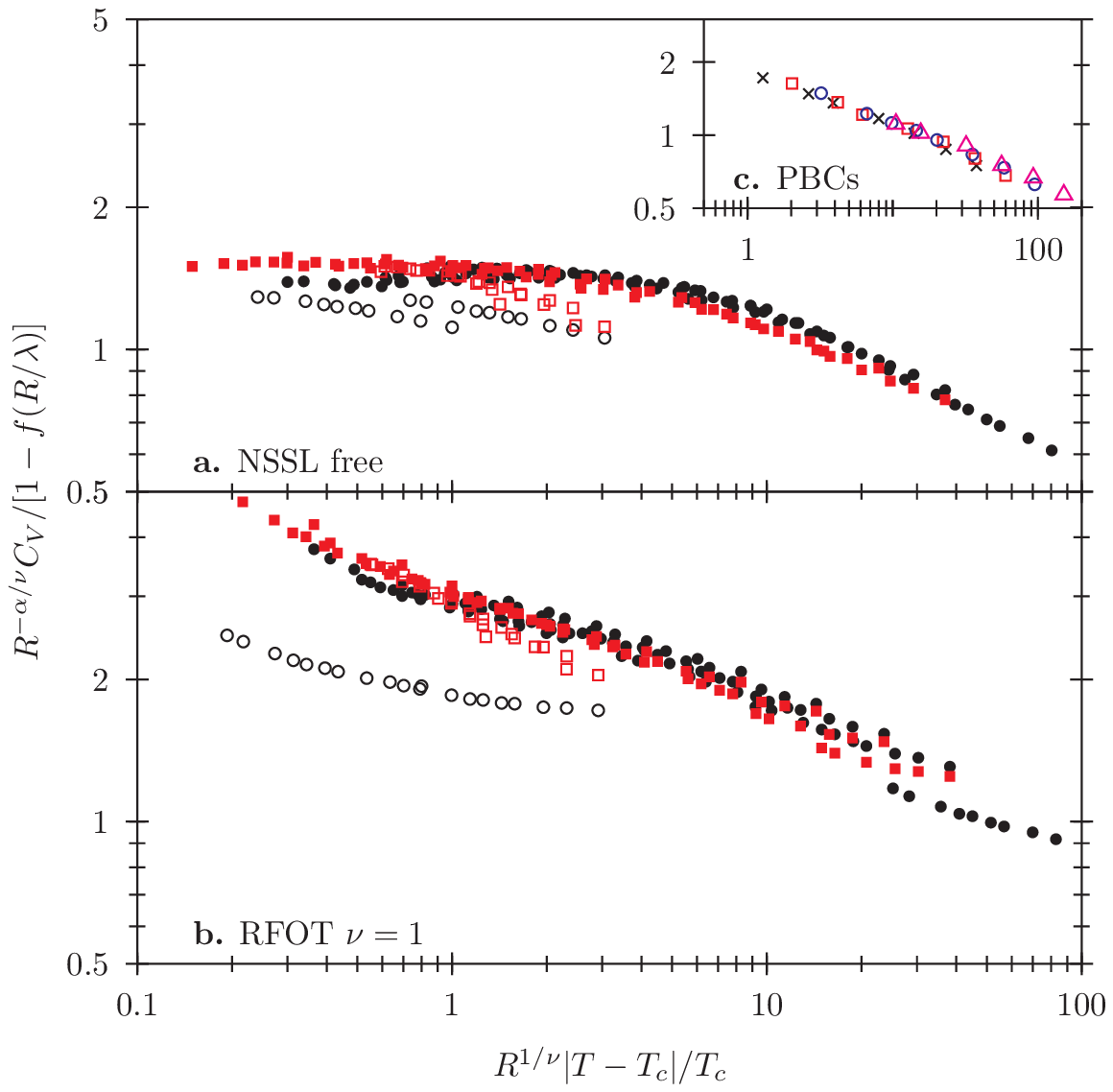}
 \caption{Scaling plots for $\Cv$ data under \acf{RBC} (red squares)
   and \acf{ABC} (black circles) according to the \ac{NSSL} (panel a)
   and \ac{RFOT} (panel b) scenarios.  $T_c$, $\alpha$, $\nu$ and
   $\lambda(T)$ are the same for both sets of data and are those
   adjusted for \ac{ABC} and quoted in Fig.~\ref{ABCfit}.  The only
   difference is that in the \ac{RFOT} case, we have set $a'=a$ in the
   $\lambda(T)$ expression to scale \ac{RBC} data.  Panel c: \ac{PBC}
   data for cubic systems of side $L=8$ (black crosses), 12.7 (red
   squares), 20.16 (blue circles), and 32 (purple triangles), scaled
   with the same exponents and $T_c$ of the \ac{NSSL} case (panel a).}
  \label{RBC}
\end{figure}

Though \ac{ABC} differ from more usual BCs such as \ac{PBC} in that
they bring forward the existence of two lengthscales, the critical
temperature and exponents are independent of the boundary conditions,
since in the $R\to\infty$ limit all observables are independent of the
boundary conditions \footnote[3]{It is understood that crystalline
  boundaries are excluded.  This is necessary to remain in the
  supercooled liquid (a crystalline border would make the whole cavity
  crystallize).  We mean independent from different BCs (such as
  \ac{ABC}, \ac{PBC} or \ac{RBC}) within the (metastable)liquid phase.
  See SM for a more complete discussion.}  unless control parameters
are such that the system is below a thermodynamic transition
\cite{book:parisi98}.  It is not possible to perform the same analysis
under \ac{PBC} in this system, because systems very small or below
$T\approx0.2$ crystallize before $\Cv$ can be measured.  The values we
have been able to obtain are compatible with the \ac{NSSL} scaling
(Fig.~\ref{RBC}c), but also with other values (see SM \footnotemark[2]
for more details).  It is not possible to collapse the \ac{PBC} data
using Eqs.~\ref{proposal} and~\ref{eqf}, since $f(x)$ is constructed
specifically for cavities (i.e.\ frozen boundaries).  We do not delve
into how the existence of two lengthscales as proposed by \ac{RFOT}
should manifest itself under \ac{PBC}; we merely point out that these
data do not contradict a scaling with a nonzero critical temperature.

Finally, we have tried a different BC on a cavity, repeating the
analysis with \ac{RBC}.  \ac{RBC} are the same as \ac{ABC} except that
the outer (fixed) particles are at random positions.  Fig.~\ref{RBC}
shows the result of applying the two most successful scalings to the
\ac{RBC} data.  \emph{This figure introduces no new parameters:}
\ac{RBC} data are scaled using the same $T_c$, $\lambda(T)$ and
exponents adjusted for the \ac{ABC} case.  Above $T_c$ both sets of
data can be scaled with the same parameters, and, at least under
\ac{RFOT}, with the same scaling function.  Below $T_c$ (open
symbols), the scaling function seems to depend on the boundaries; we
note in particular that the \ac{RBC} data can be scaled in the
\ac{RFOT} scenario without adjusting the prefactor of the $\lambda(T)$
power law.

In summary, we have studied spherical cavities with amorphous and
random BCs.  Together with swap MC that can equilibrate small
cavities, this has allowed us to do a \ac{FSS} analysis of the
specific heat, which is impossible under \ac{PBC} due to
crystallization.  We have found a peak in $\Cv$, which can be scaled
under two different scenarios, \ac{NSSL} and \ac{RFOT}.  The first
implies a divergence of $\Cv$ in the thermodynamic limit, while the
second predicts a discontinuity.  Both are of comparable quality, but
\ac{NSSL} has three free parameters, compared to one (two below $T_c$)
for \ac{RFOT}.  Within \ac{RFOT}, only $\nu=1$ gives a reasonable
collapse, suggesting a stiffness exponent $\theta=d-1$.

We finally emphasize that in all cases our collapse attempts
\emph{yield a finite $T_c,$} of around 0.17.  This means that the peak
survives the thermodynamic limit.  Since in this limit observables
must be independent of the boundary conditions (unless there is phase
coexistence \footnotemark[3]), this result implies the existence of an
anomaly or phase transition in the $R\to\infty$ limit, independently
of the particular BCs we have employed.
Investigation of more realistic and
  better glassforming liquids is needed.  This will be a
  challenge, as swap \ac{MC} is unsuitable for most systems.
  Nevertheless, these results seem a strong support for thermodynamic
theories of the glass transition.

\paragraph{Acknowledgments.}  We thank Massimiliano Viale for
technical support, and Giulio Biroli, Chiara Cammarota, Patricia
Gim\'enez, and Giorgio Parisi for discussions and suggestions.  DAM
was supported by a grant from Fundaci\'on Bunge y Born (Argentina).
TSG acknowledges support from CONICET, ANPCyT and UNLP (Argentina),
and thanks the Initiative for the Theoretical Sciences, City
University of New York, and Istituto Sistemi Complessi (Rome, Italy)
for hospitality.

\bibliographystyle{apsrev4-1}
\bibliography{lobsterALL}

\newpage
\newpage
\huge
\begin{center}
\textbf{Supplemental Material} 
\end{center}

\normalsize

\acrodef{ABC}[ABCs]{amorphous boundary conditions}
\acrodef{PBC}[PBCs]{periodic boundary conditions}
\acrodef{RBC}[RBCs]{random boundary conditions}
\acrodef{RFOT}{random first-order theory}
\acrodef{FSS}{finite-size scaling}
\acrodef{NSSL}{Not-so-simple liquid}

\def\CPBC{C_{V,PBC}}
\def\Cv{C_V}

\appendix

\section{Simulation details}

We have used the soft-sphere binary mixture of
ref.~\cite{soft-spheres:bernu87} at unit density, with size ratio 1.2,
and a smooth cut-off (details as in ref.~\cite{self:nphys08}).  The
numerical work involves two stages: preparation of the cavity and
simulation of the cavity with the frozen environment.

For the \acf{ABC} runs, the frozen border must be taken from an
equilibrated configuration at the same temperature at which the cavity
will be run.  To obtain these configurations we ran systems of
$N=8192$ soft spheres with \ac{PBC} using swap Monte Carlo
\cite{self:pre01} for temperatures $T= 1$, 
$0.683$, $0.482$, $0.350$, $0.301$, $0.260$, $0.232$, $0.214$,
$0.203$, 
$0.153$,
and
$0.107$.  %
At each temperature, 8 to 16 samples were simulated until
equilibration.  For \ac{RBC}, we simply generated 16 random
configurations.  Additionally \ac{PBC} systems of size $N=512$,
$N=2048$, and $N=32768$ were equilibrated for high ($T\geq 0.203$)
temperatures, which were later used to obtain the \ac{PBC} $\Cv$ data.

From the equilibrated or random configurations, cavities were
generated by adding to the periodic system a hard wall of spherical
shape such that density and composition within the wall are identical
to those of the \ac{PBC} system.  Then the cavity runs were carried
out, using swap Monte Carlo as before but keeping the positions of the
particles outside the wall unchanged.  We used walls of radii %
$R=1.88$, $2.12$, $2.67$, $3.06$, $3.37$, $3.86$, $4.24$, $5.23$,
$6.20$, and $7.26$, %
corresponding to cavities containing from $M=28$ to 2000 particles.
The cavities were then run, in the case of \ac{ABC} at the
temperatures listed above, and in the case of \ac{RBC} at those
temperatures plus $T=0.183$, $0.173$, $0.135$, and $0.120$.

The cavities were equilibrated and $C_V$ was measured from the energy
fluctuations:
\begin{equation}
  \label{eq:3}
  \Cv= \overline{[\langle E^2\rangle -\langle
    E\rangle^2]}/(MT^2),
\end{equation}
where $E$ is the energy and $M$ is the number of mobile (cavity)
particles, units are such that Boltzmann's constant is unity, and the
overline means average over different realizations of the boundary
(fixed particles).  In \ac{ABC} we cannot compute $\Cv$ as $d
\overline{E}/dT$, because the derivative and the average do not
commute (in fact this expression gives just the bulk $\Cv$
\cite{self:jstatmech10, mosaic:zarinelli10,
  confinement:krakoviack10}).

We have used 8 to 40 samples per radius and temperature, with sample
error ${\Delta Cv \over Cv} \lesssim 0.025$ for \ac{ABC} and ${\Delta
  Cv \over Cv} \lesssim 0.1$ for \ac{RBC} and (high temperature)
\ac{PBC}.

Simulations were performed in a 192 core Xeon E5506 2.13Ghz cluster,
devoted for this project for about one year.

\section{Equilibration}

For all our runs, we computed the (connected) energy autocorrelation
function,
\begin{equation}
  \label{eq:1}
  C(t)= \left\langle \left[ E(0)- \langle E \rangle\right] 
        \left[ E(t) -  \langle E \rangle \right] \right \rangle.
\end{equation}
We estimated the relaxation time as time $\tau$ such that
$C(\tau)=C(0)/10$, and we let the simulations run for at least
$25\tau$ before starting the data collection stage, which lasted at
least $100\tau$.  This procedure was performed self-consistently, in
the sense that the runs used to estimate $\tau$ were themselves at
least $100\tau$ long.  To compute $\Cv$ we used runs longer than
$500\tau$; for \ac{ABC} cavities, the runs were more than $1000\tau$
long.

\begin{figure}
  \includegraphics[width=\columnwidth]{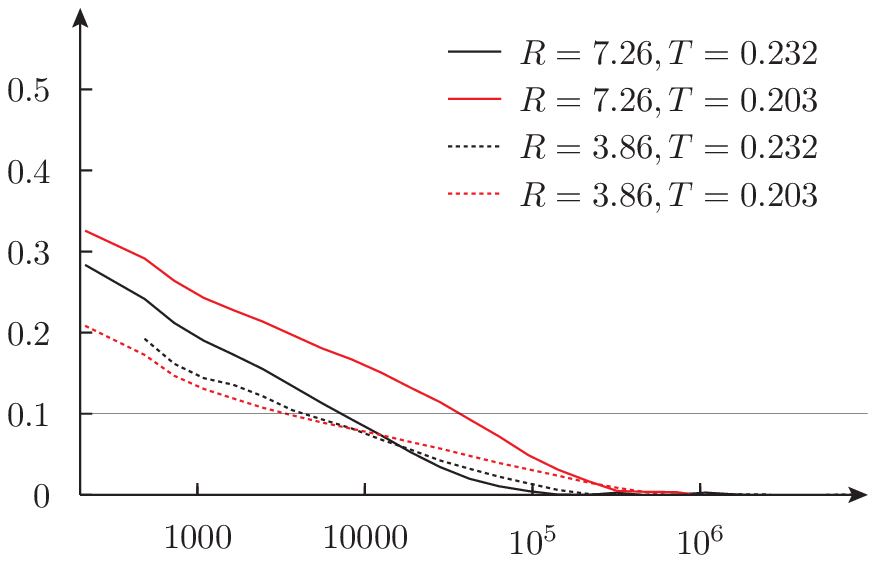}
  \caption{Representative energy autocorrelation function vs.\ time
    for two \ac{ABC} cavities at two temperatures}
  \label{fig:corrfun}
\end{figure}

\begin{figure}
  \includegraphics[width=\columnwidth]{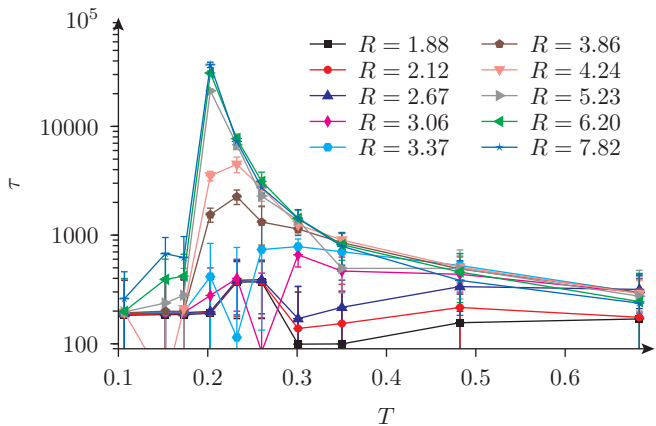}
  \caption{Correlation times for \ac{ABC}, computed as the time when
    the correlation decays to 10\% of its initial value.  These data
    include only noncrystallised samples.}
  \label{fig:taurelax}
\end{figure}

To ensure that $\tau$ was not underestimated due to the finite length
of the time series, we required that $\lvert C(t) \rvert<0.03$ for at
least 80\% of the time values.  We show four representative
autocorrelation functions in Fig.~\ref{fig:corrfun}, and the estimated
values of $\tau$ for all our \ac{ABC} cavities in
Fig.~\ref{fig:taurelax}.  One should keep in mind that we are using
swap Monte Carlo.  This is especially important for the cavity runs,
because with standard Monte Carlo the relaxation time increases
quickly as cavity size is reduced \cite{correlation-length:hocky12},
while relaxation time \emph{decreases} for smaller cavities when using
swap moves, as was found in ref.~\onlinecite{self:jcp12} looking at
overlap autocorrelation and is clear from the present data
(Figs.~\ref{fig:corrfun} and~\ref{fig:taurelax}).  Also, the
correlation times display a peak qualitatively similar to that of
$\Cv$.  The significance of this, however, is not immediately obvious,
since we are employing an unrealistic dynamics.  A further analysis of
these data will be published elsewhere.


\section{Crystallization}

For a stable thermodynamic phase, the above conditions can in
principle be fulfilled for each and every sample, given long enough
simulation time.  However, we are simulating a metastable phase (the
supercooled liquid) in finite dimension, which thus has a finite
lifetime.  Accordingly, some of the samples have to be discarded
because signs of crystallization appear before the requirements we
imposed can be fulfilled.

Supercritical crystal nuclei appeared before equilibration or before
measurement of $\Cv$ could be completed in a temperature-dependent
fraction of the samples.  This was manifested in jumps or drifts of
the energy and/or negative correlation values at long times.  Once a
supercritical nucleus appears, the ensuing coarsening process may last
for a long time.  Thus rather than attempting to bring all samples to
equilibrium, we discarded samples that failed to equilibrate when most
(more than 75\%) of the samples at the same temperature had
equilibrated according to the above requirements.

Of course, it can also happen that a crystal structure develops that
is sufficiently stable to pass the above equilibration check, so for
all samples that passed our equilibration criteria we computed the
bond orientation order parameter \cite{review:tanaka12},
\begin{equation}
  \label{eq:2}
  Q_l =\sqrt{{4\pi \over 2l+1 } \sum_{m=-l}^{m=l} |\langle Q_{lm} 
  (\mathbf{r}_{ij}) \rangle|^2},
\end{equation}
where $\mathbf{r}_{ij}= \mathbf{r}_{j}- \mathbf{r}_{i}$,
$\langle\ldots \rangle$ means average over neighbouring particles
(those whose distance is less that the first minimum in the pair
correlation function), and $Q_{lm} (\mathbf{r}_{ij})= Y_{lm}
(\theta(\mathbf{r}_{ij}), \phi(\mathbf{r}_{ij}))$, with $Y_{lm}$ the
spherical harmonics.  We have considered $Q_6$, which takes the value
of $0.3535$ for SC crystals, $0.5745$ for FCC, and intermediate values
for other relevant crystal structures (such as BC and
HCP)~\cite{liquid-thermodynamics:steinhardt83}, while for a random
configuration, $Q_6 = Q_6^\text{(ran)}=1/\sqrt{N_\text{bonds}}\simeq
1/\sqrt{6N}$ \cite{structure:errington03}.  In addition to the
equilibration requirements, we have rejected all samples with $\langle
Q_6\rangle>5Q_6^\text{(ran)}$.  In the worst cases (for \ac{PBC} at
$T=0.203$, and \ac{RBC} for the two largest radii at $T=0.120$ and
$T=0.107$), half of the samples failed one of the tests and were
excluded from the study.  For $T>0.2325$ all the samples passed the
tests, in other cases about 25\% of the samples had to be discarded.

\section{Finite-size and border effects}

When using \ac{PBC}, finite-size effects are taken into account by
introducing a scaling function $\tilde c(y)$, which depends on the
ratio of size to correlation length, $R/\xi$, and is unspecified
except for its limiting behavior.  The scaling variable is then
$y=(R/\xi)^{1/\nu}=R^{1/\nu}\lvert T-T_c\rvert/T_c$, and the scaling
prediction can be tested by plotting the data against $y$.  In our
cavity case, however, we have \emph{two} lengthscales, $\xi$ and
$\lambda$, so we expect
\begin{equation}
  \label{eq:4}
  \Cv(R,T) = \tilde g(R/\xi(T),R/\lambda(T)).
\end{equation}
Having two scaling variables means that we have not reduced the number
of variables, and thus cannot easily check our scaling by plotting
against the scaling variables.  We have thus to specify explicitly the
dependence on one of them so that we can do our scaling analysis.  In
the main text we have accordingly proposed
\begin{equation}
  \label{eq:5}
  \Cv(R,T) = R^{\alpha/\nu} \tilde c(y) [1-f(R/\lambda(T))].
\end{equation}
This is clearly an approximation, motivated by a microscopic model.
It suffers of several shortcomings, but we have stuck to it because it
introduces the least number of unknown parameters.  We discuss below
the microscopic model that leads to Eq.~\ref{eq:5}, its limitations, and
why these do not matter in the critical region, which has finally led
us to use it despite those limitations.

\paragraph{The function $f(R/\lambda)$.}

Consider the specific heat (per unit volume) of a very small volume at
distance $r$ from the center of a cavity of radius $R$, $c(r,R)$ such
that the penetration length is large.  Since the outside particles are
frozen, precisely at the border $c(R,R)$ will be very small or zero.
At a microscopic distance $\sigma$ away from the border,
configurational rearrengements will still be blocked by the surface
field produced by the outside particles, but we expect vibrations to
be possible, leading to $c(R-\sigma,R)=C_0\approx 3/2$.  Further away,
we assume an exponential decay towards the \ac{PBC} (border-free)
value $C_P$.  Hence we propose
\begin{equation}
  \label{eq:6}
  c(r,R) = \begin{cases}
    C_P + \Delta\Cv e^{-(R-\sigma-r)/\lambda}, & r<R-\sigma,\\
    C_0 (R-r)/\sigma, & R-\sigma<r\le R
  \end{cases}
\end{equation}
where $\Delta\Cv=C_0-C_P$.

The specific heat of the cavity is then obtained by integrating over
the whole sphere, $\Cv(R)=(3/R^3)\int_0^R\!\!dr\,r^2 c(r,R)$,
resulting in
\begin{equation}
  \label{eq:8}
  \begin{split}
    \Cv(R,T) ={}& (1-\sigma/R)^3C_P
    \left\{1-f\left[(R-\sigma)/\lambda\right] \right\} +
    \\
    & (1-\sigma/R)^3 C_0 f\left[(R-\sigma)/\lambda\right] +
    h(\sigma/R),
  \end{split}
\end{equation}
where $f(x)$ is given by Eq.~2 of the main text and the last term is a
microscopic contribution
\begin{equation}
  \label{eq:9}
  h(x) = 3C_0 x \left(\frac{1}{2}-\frac{2x}{3}+\frac{x^2}{4} \right).
\end{equation}
Finite-size, non-border effects are then taken into account by writing
$C_P=R^{\alpha/\nu}\tilde c(y)$.  When $\sigma=0$ and $C_0=0$,
Eq.~\ref{eq:8} reduces to Eq.~\ref{eq:5}.

For $R\gg\sigma$ and $\lambda\to\infty$, Eq.~\ref{eq:8} gives
$\Cv\to C_0$.  Thus it seems reasonable to keep $C_0\neq0$, unlike
what we have done.  One can indeed do so, but it turns out that $C_0$
and $\sigma$ cannot be arbitrary: for a given $\sigma$, only one value
of $C_0$ will make the curves collapse.  In particular, for
$\sigma=0$, this value is $C_0=0$.  This relationship can be better
understood through the following considerations, which show that the
value $C_0$ actually never plays the role of an observable quantity.

\paragraph{The critical region.}

The critical region is defined by a finite value of $y$ ($y_0$ say),
so that in this region $R\sim\xi$.  This is where the transition
should take place, because the correlation length is of the order of
the system size.  When there are no border effects, the peak (i.e.\
the finite-size transition) is readily found to occur at
$T_p= T_c + y_0 T_c/(R^{1/\nu})$, where $y_0$ is the position of the
maximum of $\tilde c(y)$.  Including the border contributions in
Eq.~\ref{eq:8}, the condition for $\Cv$ to have a maximum (at fixed
$R$) is
\begin{equation}
  \label{eq:11}
  \tilde c'(y) = \frac{c(y)-R^{-\alpha/\nu}C_0}{1-f(x)} \frac{f'(x) R^{-1/\nu}
    (R-\sigma)}{\lambda^2}\frac{\partial \lambda}{\partial T},
\end{equation}
with $x=(R-\sigma)/\lambda$.  Since $f'(x)\sim x^{-2}$,
$\partial \lambda/\partial T \sim \lambda^3$, and
$\lambda\sim \lvert T-T_c\rvert^{1/2} \sim \xi^{1/2\nu}$, the r.h.s.\
is of order $R^{-1/\nu}\lambda^3/R \sim R^{1/2\nu-1}$, so that
self-consistently near the peak at large $R$ the maximum is given by a
finite value of $y$ near $y_0$, and $R\sim\xi$.

\paragraph{The limit $R/\lambda\to0$.}

In \ac{RFOT} the length $\lambda$ diverges as $T\to T_c$, as does
$\xi$, but with a smaller exponent, so that $\xi\gg\lambda$.  It
follows that the limiting value $f_0\equiv \lim_{x\to0}f(x)$ is
irrelevant in the critical region, because there
$R/\lambda \sim \xi/\lambda \gg1$.

The value of $f_0$ plays a role at $T_c$ (i.e.\ \emph{outside} the
critical region for finite samples), as can be seen inverting
Eq.~\ref{eq:8}:
\begin{equation}
  \label{eq:10}
  \tilde c(y) = R^{-\alpha/\nu}
  \frac{\left[\Cv(R,T)-h(\frac\sigma R)\right] (1-\frac\sigma R)^{-3} - C_0
    f[\frac{R-\sigma}\lambda]} {1-f[\frac{R-\sigma}\lambda]}.
\end{equation}
Since $\lambda\to\infty$ at $T_c$ and $f_0=1$ for our $f(x)$,
$\tilde c(y)$ will develop a divergence for $y\to0$ unless a delicate
cancellation occurs in the numerator.  In particular, when
$C_0=\sigma=0$, there is no cancellation and $\tilde c(y\to0)$
diverges.  This is inconsistent with the expectation that
$\tilde c(y)$ is the \ac{PBC} scaling function, which is defined to be
finite at $y=0$.  The origin of this problem is explained next.

\paragraph{Diverging $\lambda$ and finite cavities.}

\begin{figure}
  \includegraphics[width=\columnwidth]{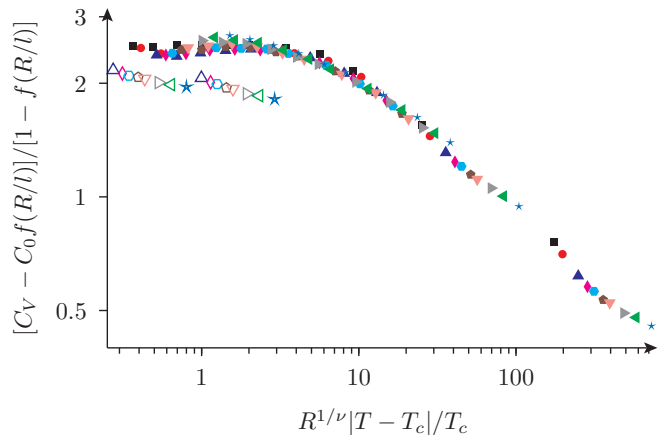}
  \caption{Scaling of \ac{ABC} data using Eqs.~\ref{eq:8} and
    $l(x)=1/(D_0+x)$.  Parameters are $\nu=1$, $\alpha=0$, $Tc=0.175$,
    $C_0=0$, $\sigma=0.2$, $D_0=20$.}
  \label{fig:nondiverging-lambda}
\end{figure}

The basic idea of \ac{FSS} is that $\xi$ in a finite system is never
infinite, but is cut-off by the system size $R$.  This is how the
properties of the scaling function are inferred \cite{book:newman99}.
Clearly, the same applies to $\lambda$.  However, our simple model for
the border effects does not incorporate this: that is the reason for
the divergence of the scaling function at small $y$.  The function
$\tilde c(y)$ diverges to compensate for the unphysical feature of an
infinite $\lambda$ at $T_c$ and finite $R$.  One way to cure this
would be to replace $\lambda$ by an effective length $l(R/\lambda)$,
given by
\begin{equation}
  \label{eq:7}
  l(R/\lambda) = R \tilde l(R/\lambda), 
  \qquad \tilde l(x) =
    \begin{cases}
      \text{const}, & x\to0,\\
      x^{-1}, &x\to\infty,
    \end{cases}
\end{equation}
so that $l$ remains finite for $T\sim T_c$ when $R$ is finite.  This
can be done (see Fig.~\ref{fig:nondiverging-lambda}), but the cost is
to introduce many additional unknowns.

\paragraph{Summary.}

The value of the specific heat at the border of the cavity set by a
particular microscopic model is irrelevant in the critical region
where $R\sim\xi\gg\lambda$.  Our simple Eqs.~1 and~2 of the main text
suffer from a more drastic problem: they unphysically allow for
$\lambda$ to become actually infinite at $T=T_c$ and finite $R$.
Because of this, the microscopic details artificially show up near
$T_c$, with the result that the estimated scaling function
$\tilde c(y)$ has to compensate for this shortcoming of our treatment
of border effects, differing (even possibly diverging) from the
\ac{PBC} scaling function at small $y$.  While this problem can be
cured by introducing an effective $l(R/\lambda)$ as shown above, this
involves additional unknowns.  We have hence decided to keep unknown
constants to a minimum, at the expense of not correctly estimating
$\tilde c(y)$ for small $y$.  We repeat, though, that these issues do
not affect the critical region.

\section{Periodic boundary conditions}

Since our \ac{FSS} analysis points to a result (a finite $T_c$) that
holds in the thermodynamic limit independently of the boundary
conditions, it is natural to ask whether this result could be obtained
using the more usual \ac{PBC}.  We have shown in the main text
(Fig.~3) that \ac{PBC} data are compatible with the scaling that
yields a finite $T_c$.  Here we explain in more detail why \ac{PBC}
data are however not enough to establish such scaling.

Basically, with \ac{PBC} one can obtain much less datapoints of $\Cv$
due to crystallization.  While the spherical geometry makes
crystallisation very rare for cavities (and rarer for the smaller
ones), in the periodic geometry crystallization occurs too quickly to
obtain meaningful data for $N\lesssim 500$ particles and temperatures
below $\approx 0.26$.  In larger systems the problem is less severe,
but lower temperatures are still problematic.  While we have been able
to \emph{thermalize} 8192 particles down to $T=0.107$ as stated above,
this does not mean that $\Cv$ \emph{can be measured reliably} down to
the same temperature.  We have required a time of at least $100\tau$
to deem the system equilibrated, but the goal of a 10\% relative error
in $\Cv$ could not be reached with \ac{PBC} for $T<0.2$ due to the
limited number of samples that could be run for more than $100\tau$
without crystallization ($\Cv$ was measured in runs lasting at least
$500\tau$).  We have proceeded conservatively and considered
$\Cv$ for \ac{PBC} only for $T\ge0.2$.

\begin{figure}
  \includegraphics[width=\columnwidth]{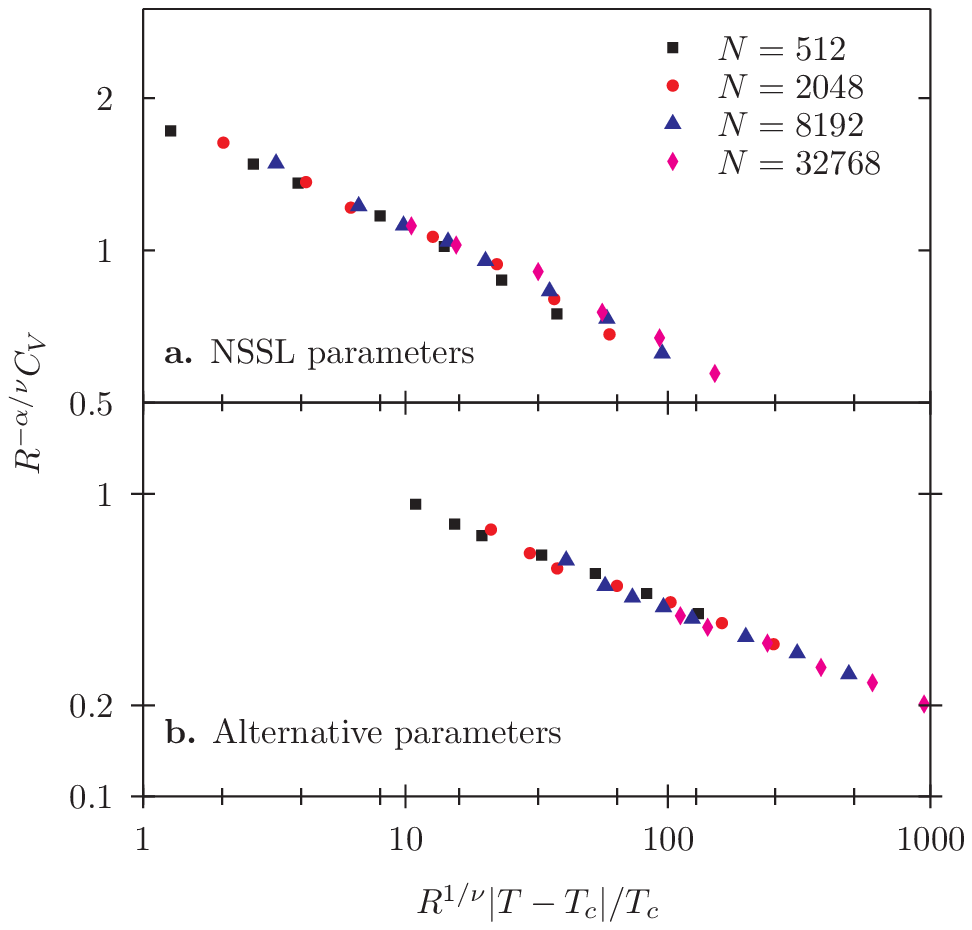} 
  \caption{\ac{PBC} specific heat scaling plots with two sets of
    parameters, showing that the scaling is marginal.  Top: parameters
    as in the NSSL scaling (main text), $\nu=1$, $\alpha=0.2$,
    $T_c=0.175$.  Bottom: $\nu=0.7$, $\alpha=0.35$, $T_c=0.13$.}
  \label{fig:scaling-marginal}
\end{figure}

With these data it is not possible to obtain reliable values for
critical temperature and exponents because the scaling is marginal,
allowing many different values (Fig.~\ref{fig:scaling-marginal}).
This is why cavities are needed to perform the \ac{FSS} analysis.

\section{Phase transition and dependence on boundary conditions}

\label{sec:boundary-conditions}

We have argued that since in the thermodynamic limit observables must
be independent of the boundary conditions unless the system is below a
phase transition, our results favor a scenario with a thermodynamic
glass transition independently of the particular boundary conditions
we have employed.  Since our system \emph{is} already below a phase
transition, we expand the argument a bit more, to show how it can
safely be extended to a metastable phase.

When a Kauzmann-like transition (i.e.\ from the supercooled liquid to
a phase different from both liquid and crystal) is discussed, it is
understood that it involves two \emph{metastable} phases: the liquid and
the ideal glass (or whatever one chooses to call it) are both
metastable with respect to the crystal.  However, the crystal is
usually ignored, implying that one can prove or assume that the time
for crystallisation is long enough to allow for the metastable phases
to equilibrate before crystallisation begins.  Here, we have similarly
ignored not only the crystal that eventually forms for very large
times, but also the crystalline boundary conditions that would not
only select particular crystalline forms, but also greatly accelerate
crystallisation.

Recall that boundary conditions are important in the thermodynamic
limit when the system is in a broken symmetry phase, where there is
coexistence among the different broken-symmetry states (which are
themselves connected by an operation of the symmetry that has been
broken).  To clarify, consider the Ising model below $T_c$: the
up-down symmetry is broken.  This means that the equilibrium state can
have positive or negative magnetisation, and that with \ac{PBC} both
of them will be reached with equal probability.  On the other hand,
introducing walls with positive (negative) magnetisation will select
the state with positive (negative) magnetisation, even in the limit of
infinite systems.  The phases that coexist are not the paramagnet and
the ferromagnet, but the ferromagnetic phases with positive and
negative magnetisation.  Dependence on the boundary conditions means
that there exist boundaries that can select the final state, but
there are also boundaries (like \ac{PBC} or \ac{RBC}) that don't
matter much: the final magnetisation will still be random.

In our case of a liquid below the melting transition, the coexisting
phases are the different possible crystals (orientations, shifts, and
in general the crystal symmetry operations).  A cavity with a
crystalline wall would crystallise in the crystal chosen by the wall.
But amorphous (i.e. liquid) boundaries are not selecting a particular
crystal (rather, they inhibit, though not completely suppress, the
crystal).

What we ask is whether the metastable liquid can transition to another
phase (also metastable with respect to the solid), and the boundaries
we use are noncrystalline: they could select different amorphous
states, but with respect to the crystal they are mostly irrelevant (as
the \ac{RBC} in the Ising example above): for all our boundary
conditions, a large enough system crystallises in the long run.  If we
ignore the crystal (actually, if we are careful to study the system up
to the appearance of crystal nuclei, as we do), then we can ask
whether there is a symmetry breaking \emph{within} the metastable
state (a metastable liquid-metastable glass transition).  Above this
transition, the thermodynamic limit must be independent of the
boundary conditions, up to the appearance of the crystal (which would
be much accelerated if one chose a crystalline border).  Below $T_c$,
the coexisting phases would be e.g.\ the different states of \ac{RFOT}
(that break the replica symmetry).  Our boundary conditions then might
select one of those states below $T_c$, leading to a dependence on the
amorphous boundaries which is not present above $T_c$.  Again, this
ignores a possible crystalline boundary, which would have the same
effect above and below $T_c$.

\end{document}